\documentclass[fleqn,usenatbib]{mnras}

\usepackage{newtxtext,newtxmath}

\usepackage[T1]{fontenc}
\usepackage{ae,aecompl}

\usepackage{graphicx}	
\usepackage{amsmath}	
\usepackage{amssymb}	

\title[RGZ: New GRGs in the RGZ DR1 catalogue]{Radio Galaxy Zoo: New Giant Radio Galaxies in the RGZ DR1 catalogue}

\author[H Tang et al.]{
H.~Tang$^{1}$,\thanks{E-mail: hongming.tang@manchester.ac.uk}
A.~M.~M.~Scaife$^{1,2}$,
O.~I.~Wong$^{3,10,11}$,
A.~D.~Kapi\'{n}ska$^{4}$,
L.~Rudnick$^{5}$,
\newauthor
S.~S.~Shabala$^{6}$,
N.~Seymour${^7}$, 
R.~P.~Norris$^{8,9}$,
\\ \\ \\
$^{1}$Jodrell Bank Centre for Astrophysics, University of Manchester, Manchester M13 9PL, UK\\
$^{2}$The Alan Turing Institute, Euston Road, London, NW1 2DB, UK\\
$^{3}$ICRAS-M468, University of Western Australia, Crawley, WA 6009, Australia\\
$^{4}$ National Radio Observatory Astronomy (NRAO), POBox 0, Socorro NM, USA\\
$^{5}$School of Physics and Astronomy, University of Minnesota, 116 Church St. SE, Minneapolis, MN 55455, USA\\
$^{6}$School of Natural Sciences, Private Bag 37, University of Tasmania, Hobart, TAS 7001, Australia\\
$^{7}$International Centre for Radio Astronomy Research, Curtin University, Perth, Australia\\
$^{8}$CSIRO Astronomy and Space Science, Australia Telescope National Facility, PO Box 76, Epping, NSW 1710, Australia\\
$^{9}$Western Sydney University, Locked Bag 1797, Penrith, NSW 2751, Australia\\
$^{10}$CSIRO Astronomy and Space Science, PO Box 1130, Bentley, WA 6102, Australia\\
$^{11}$ARC Centre of Excellence for All Sky Astrophysics in 3 Dimensions (ASTRO 3D), Australia\\
\\ \\ 
}

\date{Accepted XXX. Received YYY; in original form ZZZ}

\pubyear{2019}

\begin{document}
\label{firstpage}
\pagerange{\pageref{firstpage}--\pageref{lastpage}}
\maketitle

\begin{abstract}
In this paper we present the identification of five previously unknown giant radio galaxies (GRGs) using Data Release 1 of the Radio Galaxy Zoo citizen science project and a selection method appropriate to the training and validation of deep learning algorithms for new radio surveys. We associate one of these new GRGs with the brightest cluster galaxy (BCG) in the galaxy cluster GMBCG\,J251.67741+36.45295 and use literature data to identify a further 13 previously known GRGs as BCG candidates, increasing the number of known BCG GRGs by $>60\%$. By examining local galaxy number densities for the number of all known BCG GRGs, we suggest that the existence of this growing number implies that GRGs are able to reside in the centres of rich ($\sim 10^{14}$\,M$_{\odot}$) galaxy clusters and challenges the hypothesis that GRGs grow to such sizes only in locally under-dense environments.
\end{abstract}




\section{Introduction}
Giant Radio Galaxies (GRGs) are the largest radio galaxies in the Universe. Originally defined to be those radio galaxies with projected linear sizes greater than 1\,Mpc, in a cosmology with $\rm H_{\rm 0} = 50~km~s^{-1}~Mpc^{-1}$ \citep{Willis1974}, the GRG size limit is now equivalent to 700\,kpc in a $\rm \Lambda$CDM cosmology with the Planck 2016 parameters \citep{Planck2016,Dabhade2019}. It is thought that GRG sizes might be caused by high kinetic jet power \citep{Wiita1989} and it has been shown that the size of a radio source is positively correlated with source radio luminosity and jet power \citep{Shabala2013}. Alternatively, it has also been proposed that the gigantic size of GRGs might be caused by the comparative longevity of their jets \citep{Subrahmanyan1996}, or due to the radio source growing in a low density environment \citep{Malarecki2015}.

The role of local environment in GRG formation was first considered by \cite{Ishwara1999} who compared 53 GRGs in the literature with 3CR radio sources \citep{Laing1983} of smaller sizes. \cite{Ishwara1999} found that GRGs share a marginally higher separation ratio of hotspot distances from the nucleus, which \cite{Ishwara1999} suggest might be caused by the interaction of energy carrying beams and cluster-sized density gradients far from the source host galaxy. This has in turn led to GRGs being used as probes of the low ambient density warm-hot intergalactic medium \citep[WHIM;][]{Safouris2009,Peng2015}.

Another environmental consideration is the local galaxy density around GRGs. GRGs have typically been found in under-dense environments, and it has been proposed that such reduced galaxy densities facilitate these radio galaxies to grow larger \citep{Malarecki2015}. However, a number of other studies have found that there is no correlation between radio source linear size and local galaxy density \citep{Komberg2009,Kuzmicz2018,Ortega2018}. Moreover, the recent discovery of more than 20 GRGs that
not only reside in galaxy cluster environments as found in \citet{Seymour2020}, but are also the brightest galaxy in these clusters (brightest cluster galaxies; BCGs) has also challenged this hypothesis \citep{Dabhade2017, Dabhade2019}.

From a galaxy evolution perspective, GRGs represent the tail of the radio galaxy size distribution. A comprehensive study of the shape of this distribution requires consistent sampling of both GRGs and smaller radio galaxies. However, traditional methods of cross-matching large scale radio surveys, like the Faint Images of the Radio Sky at Twenty-Centimeters \citep[FIRST;][]{Becker1995} survey, with optical/infrared surveys such as those obtained using the Wide-field Infrared Survey Explorer \citep[\textit{WISE};][]{Wright2010}, e.g. the AllWISE image atlas and catalogue \citep{Cutri2013}, are complicated by scale-dependent observational selection effects, as well as the uncertainties in cross-matching which arise when dealing with diffuse or complex radio emission. 

The physical size of a source can be calculated if its host galaxy redshift ($z$) and its largest angular size (LAS) are available. These require validated cross-identification of radio components and their host galaxy. Traditionally, a limited number of experts would first identify radio source components and then cross-match their optical/infrared hosts \citep[e.g.,][]{Subrahmanyan1996,Lara2001,Machalski2001,Schoenmakers2001,Saripalli2005,Machalski2007,Solovyov2011}. Recently, thanks to the availability of large optical and radio surveys, \citet{Dabhade2019} discovered 225 new GRGs using the Value Added Catalogue \citep[VAC;][]{Williams2019} of the LOw Frequency ARray \citep[LOFAR;][]{Harrlem2013}. Most compact sources in the VAC catalogue are selected by cross-matching the LOFAR Two-metre Sky Survey Data Release 1 catalogue \citep[LoTSS DR1;][]{Shimwell2017,Shimwell2019} with a catalogue of matches between the Panoramic Survey Telescope and Rapid Response System \citep[Pan-STARRS;][]{Kaiser2002,Kaiser2010,Chambers2016} catalogue and the AllWISE catalogue, using a likelihood ratio method \citep{Williams2019,Dabhade2019}. However, diffuse and complex sources in the catalogue are cross-matched by visual inspection using the citizen science LOFAR Galaxy Zoo project \citep[LGZ;][]{Williams2019}. Among the 231,716 sources of LoTSS DR1 that have optical/IR identifications, only 0.1\%  are found to be GRGs \citep{Williams2019,Dabhade2019}.

Citizen science offers an alternative to more traditional methods of building large cross-matched radio galaxy catalogues. Radio Galaxy Zoo \citep[RGZ;][]{Banfield2015} is an online citizen science project which aims to cross-match extended radio sources from the FIRST survey \citep{Becker1995} and the Australia Telescope Large Area Survey \citep[ATLAS;][]{Franzen2015} with their host galaxies in the infrared waveband, using data from the AllWISE survey and the SIRTF Wide-Area Infrared Extragalactic Survey \citep[SWIRE;][]{Lonsdale2003}. RGZ offers its volunteers a 3$\times$3\,arcmin$^2$ cutout from the FIRST survey with radio contours starting at 3$\rm \sigma_{rms}$ on top of a \textit{WISE} 3.4\,$\rm \mu $m image. Project participants are asked (a) to identify radio components of a source from an image, (b) to select the infrared host galaxy of the corresponding radio source, and (c) to check if there are additional sources without existing identifications present in the same image \citep{Banfield2015}. The project is intended to provide the foundation of a large cross-matched radio galaxy catalogue.

Both citizen science and traditional astronomy methods are now being used as the foundation for recent studies that employ automatic radio morphology classification using deep learning algorithms. Using a number of radio galaxy catalogues which include radio morphology classification, e.g. FRICAT; \citealt{Capetti2017a}, FRIICAT; \citealt{Capetti2017b}, the Combined NVSS-FIRST Galaxies (CoNFIG) sample; \citealt{Gendre2008,Gendre2010} and \citealt{Mingo2019}, several automatic radio morphology deep learning classifiers have been developed \citep[e.g.][]{Aniyan2017,Alhassan2018,Lukic2018,Lukic2019,Ma2019}. These automatic classifiers are built to extract morphological features from input images for classification. For general radio galaxy classification \citep[FR\,I/FR\,II;][]{FR1974}, these applications have achieved model accuracies comparable to visual inspection. However, these deep learning algorithms require their individual image inputs either to have a common input image size or to be resized to a common size \citep{Lukic2019}. Since GRG identification requires LAS estimation, training a deep learning based GRG classifier under these constraints would require an image training dataset with image sizes large enough to enable an algorithm to estimate source LAS for very extended objects. Considering the memory limits of state-of-the-art GPUs, the image sizes required for such an algorithm in the GRG case are likely to make such a general approach highly computationally expensive. Consequently, in the case where image size is restricted due to memory limitations, as well as the potential for confusion due to multiple objects in the field, careful consideration must be given to the effects of selection bias in the use of such machine learning approaches.

In this paper, we identify five new GRGs selected from the RGZ Data Release 1 (RGZ DR1; Wong et al., in prep.). RGZ DR1 is a manually cross-matched radio galaxy catalogue, using the efforts of more than 12,000 citizen scientist volunteers (Wong et al., in prep.). Unlike previous GRG identification studies, this work uses a process compatible with the constraints imposed by current deep learning algorithms.
In \S~\ref{sec2_0} we describe the initial source selection process; in \S~\ref{sec3_0} we describe the validation process for identifying GRGs; in \S~\ref{sec4_0} we draw comparisons with other GRG identification studies, including their comparative selection effects and resulting impact on deep learning algorithms. We also discuss the characteristics and environments of these newly identified GRGs in a wider context; and in \S~\ref{sec5_0} we draw our conclusions.

In this work we assume a $\rm \Lambda$CDM cosmology with $\Omega_{\rm m}$ = 0.31 and a Hubble constant of $\rm H_{\rm 0} = 67.8~km~s^{-1}~Mpc^{-1}$ \citep{Planck2016}. The AllWISE magnitudes we adopted in the work follow the Vega magnitude system \citep{Wright2010} Finally, we adopted the spectral index definition as $S \propto \nu^{\alpha}$ throughout the work, where $\alpha$ is the spectral index.
 
\section{Source selection}
\label{sec2_0}

RGZ DR1 is a catalogue from the first 2.75\,years of the RGZ project (Wong et al., in prep.). Within the catalogue, 99.2\% of classifications used radio data from the FIRST survey, and the remainder used data from the ATLAS survey. Each source classification has a user-weighted consensus fraction (consensus level) $>0.65$ (Wong et al., in prep.). The LAS of each source in the RGZ DR1 is estimated by measuring the hypotenuse of a rectangle that encompasses the entire radio source at the lowest radio contour \citep{Banfield2015}. This method is generally reliable if the radio lobes of a source are correctly identified and the source is not severely bent.

The RGZ DR1 catalogue contains information on individual radio galaxies and their associated radio components, and also a table of cross-matched host galaxies. In this study, we used the catalogue of cross-matched host galaxies as our primary input sample. The original catalogue contains $\sim$140,000 entries. From this catalogue we removed sources without FIRST data available or without host galaxy redshift data in the Sloan Digital Sky Survey \citep[SDSS;][]{Alam2015}. This reduced the sample to 11\,549 entries.

Although the RGZ DR1 catalogue requires entries to have a minimum consensus level of 0.65, we found duplicate entries that had (i) the positions of two radio sources separated by less than the pixel size of the FIRST survey (1.8\,arcsec), or (ii) multiple host galaxies identified within the same extended radio source. We identified 186 of the first instance and 147 of the second instance. We visually inspected all these source pairs. We then eliminated one source from each pair of the first instance if their LAS and host galaxy redshift were identical. In the second instance, we retained the source in each pair which had a position closer to the \textit{WISE} host galaxy position. This process removed a further 312 objects, which reduced the number of objects in the sample to 11\,237 entries. 
\begin{figure}
\setlength{\unitlength}{1cm}
\includegraphics[width=0.45\textwidth,height=0.5\textwidth]{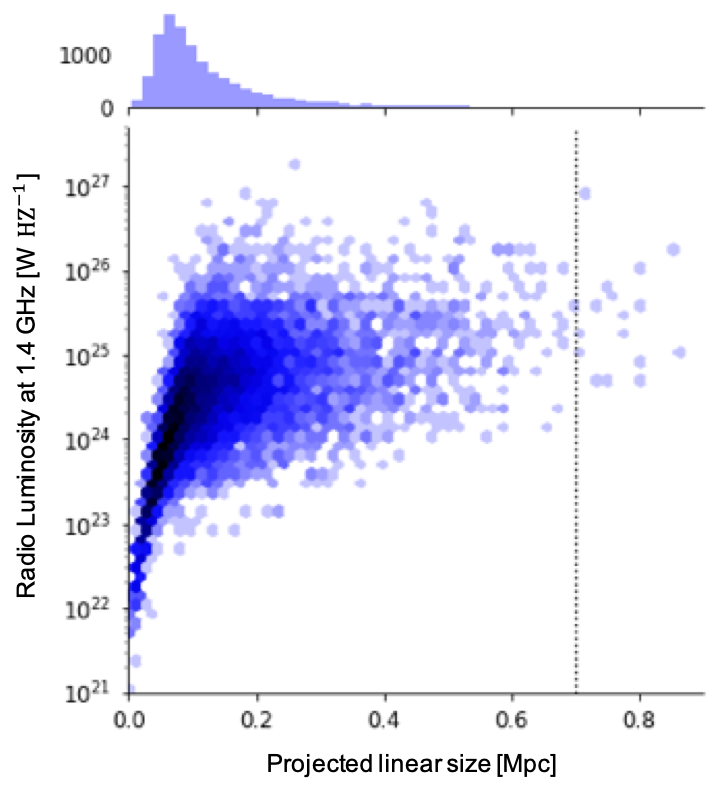}
\caption{Upper: The projected linear size histogram of the adopted 11\,237 RGZ DR1 candidates. Lower: The size-radio luminosity diagram of the same candidates. The color of each hexagon in the diagram represents the source number density with corresponding size and radio luminosity at 1.4 GHz. The dashed line refers to 700 kpc of linear size. }
\label{fig:size_lumin}
\end{figure}

Fig.~\ref{fig:size_lumin} shows the size-luminosity diagram of this sample. The projected linear size and 1.4\,GHz radio luminosity for each object were calculated using the catalogued RGZ DR1 source LAS, integrated flux density measured from FIRST images, and the host galaxy redshift, assuming a typical spectral index of $\alpha=\,$-0.7. The majority of the samples share modest radio luminosity and projected linear size. Sources with larger size tend to have higher radio luminosity. Within the sample there are 17 objects which have a projected physical size greater than 700\,kpc. The FIRST images for each of these entries were visually inspected, and one additional repeated object was identified and removed. We then crossed-matched the remaining 16 objects with the GRG catalogues of \cite{Kuzmicz2018}, \cite{Dabhade2019} and \cite{Kozie2020}, and found that \citet{Kuzmicz2018} had previously recorded three of the objects: J0929+4146, J1511+0751, and J1521+5105. We also cross-matched the recent \citet{Proctor2016} inspired GRG candidate catalogue from \citet{Dabhade2020}, and found no overlap. The remaining 13 candidate objects were not found to match any previously known GRG.

\section{Giant Radio Galaxy Identifications}
\label{sec3_0}

For the 13 candidate GRGs, we refined their identifications and measurements using manual inspection of the data. This inspection was used to clean the dataset in three steps:

\begin{enumerate}

\item We examined the relationship between the radio structure and the assigned infrared host galaxy of each entry using a \textit{WISE} 3.4\,$\mu$m image centred on the estimated central radio emission position of the radio galaxy. This process removed three objects where no clear relationship between the radio lobes and the host was seen.  

\item We re-calculated the LAS of each radio galaxy using the HEALPix Ximview software \citep[HEALPix;][]{Gorski2005} and the FIRST images, and compared these values with the LAS recorded in the RGZ DR1 catalogue. This process identified three fields containing two radio galaxies that had been misidentified as a single source. In two further fields, we found that the source LAS was overestimated due to confusion with neighbouring sources and that this had also caused the host galaxy to be misidentified. These five objects were removed from the candidate list.

\item For objects with misidentified host galaxies (point $(i)$ above), we made a renewed host galaxy search using the NASA Extragalactic Database (NED) and the SDSS Sky Server \citep[SDSS DR15;][]{Aguado2019}. The mid-point of the radio emission was chosen to be the search centre in each case. Given that each image had a side of 3\,arcmin, we searched within a radius of 1.5\,arcmin. In those cases where a host redshift was found in SDSS DR15, we re-measured the projected linear size of each radio source. This check showed that none of the misidentified sources  had a projected linear size larger than 700\,kpc.
\end{enumerate}

\begin{figure*}
\setlength{\unitlength}{1cm}
\includegraphics[width=0.8\textwidth]{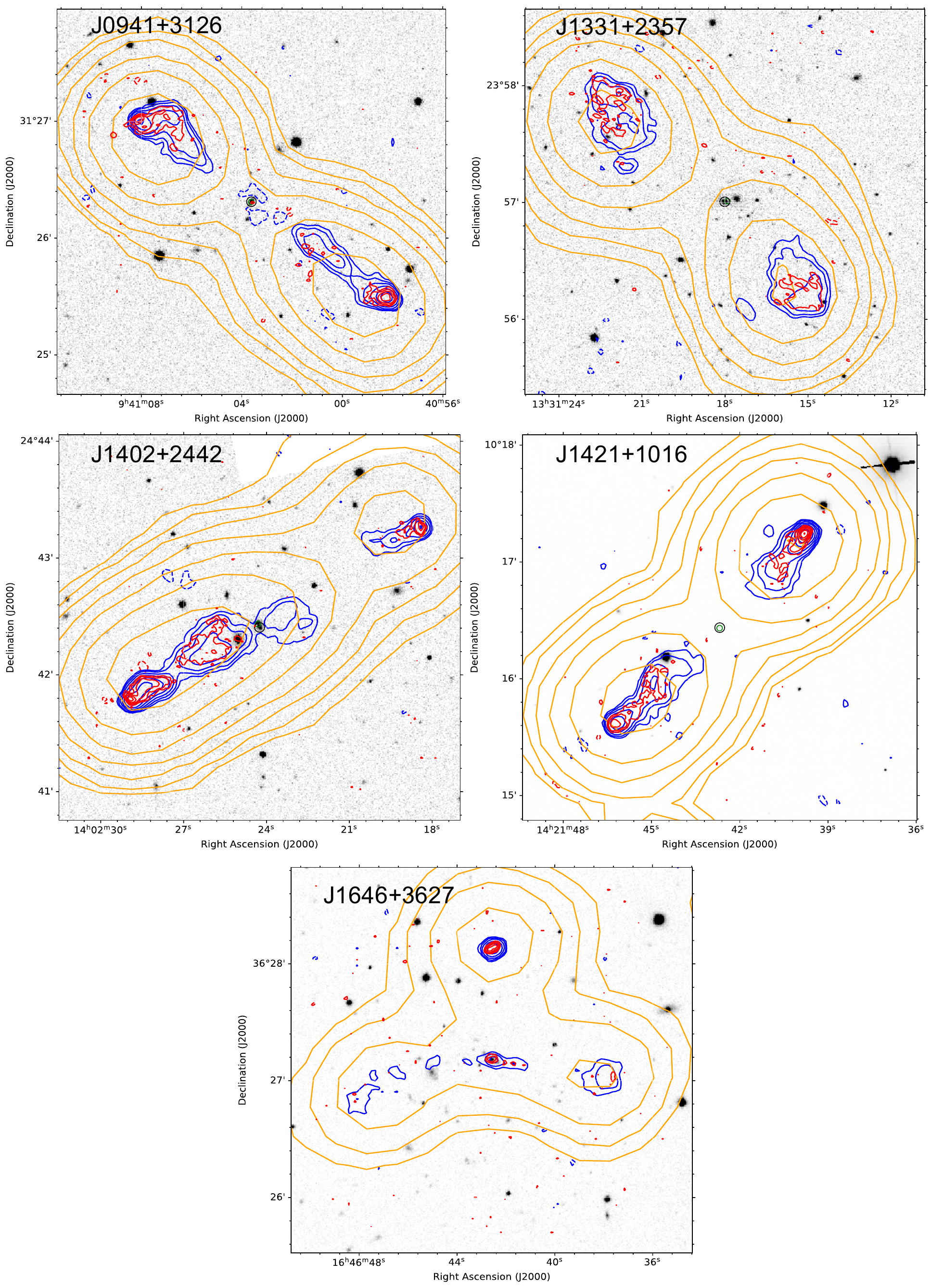}
\caption{The new GRGs identified in this work. The figure shows radio-near infrared overlays of these sources, using SDSS i-band images rather than \textit{WISE}, given their better angular resolution. The orange, blue and red radio contours for each source from the NVSS at 1.4\,GHz, FIRST at 1.4\,GHz and the Karl G. Jansky Very Large Array Survey \citep[VLASS;][]{Lacy2019} at 3\,GHz, respectively, are shown on each image from 3$\sigma_{\rm rms}$ increasing in steps of a factor of 2. The dashed lines are -3$\sigma_{\rm rms}$ of the same survey. The \textit{WISE} candidate 
host galaxy identified by RGZ DR1 is shown as a green ring, while possible SDSS host galaxies we found are shown in a black ring. The host galaxy position of J1646+3627 concides with the peak brightness of its VLASS/FIRST images. Moreover, the diffuse and compact radio emission above J1646+3627 is irrelevant to the source.} 
\label{fig:GRG_with_NVSS_contour}
\end{figure*}

This three-step data cleaning process resulted in a final sample of 5 GRGs. Fig.~\ref{fig:GRG_with_NVSS_contour} shows the images of these sources; Table~\ref{tab:GRG basic information}, Table~\ref{tab:GRG infrared properties} and Table~\ref{tab:GRG multi-survey flux summary} summarize the redshift, LAS, linear size, infrared and radio properties of each object. Redshifts in the tables are extracted from SDSS DR15. Source LASs have been manually re-measured, but are generally consistent (typically 0.5\% larger) with those from the original DR1 catalogue.

For the newly identified GRGs, we used visual inspection of the FIRST data to classify each source by morphology and found that four of the five objects to be FR\,II type. The fifth source, J1646+3627, has an ambiguous morphology. 

All five sources have comparatively high radio luminosities, with $\log P_{\rm 1.4} \,{\rm [W/Hz]} >$ 25.1, the mean total radio luminosity of FR~II objects as determined by \citet{Kozie2011}. Since the host galaxies in each case have W1-W2$< 0.8$ and W2-W3$<3.5$, where W1, W2, and W3 are the \textit{WISE} observed source magnitudes at 3.4\,$\rm \mu $m, 4.6\,$\rm \mu $m and 12\,$\rm \mu $m \citep{Cutri2013}, they are likely to be either elliptical or intermediate disk galaxies \citep{Jarrett2017}. 

\begin{figure}
\setlength{\unitlength}{1cm}
\includegraphics[width=0.45\textwidth]{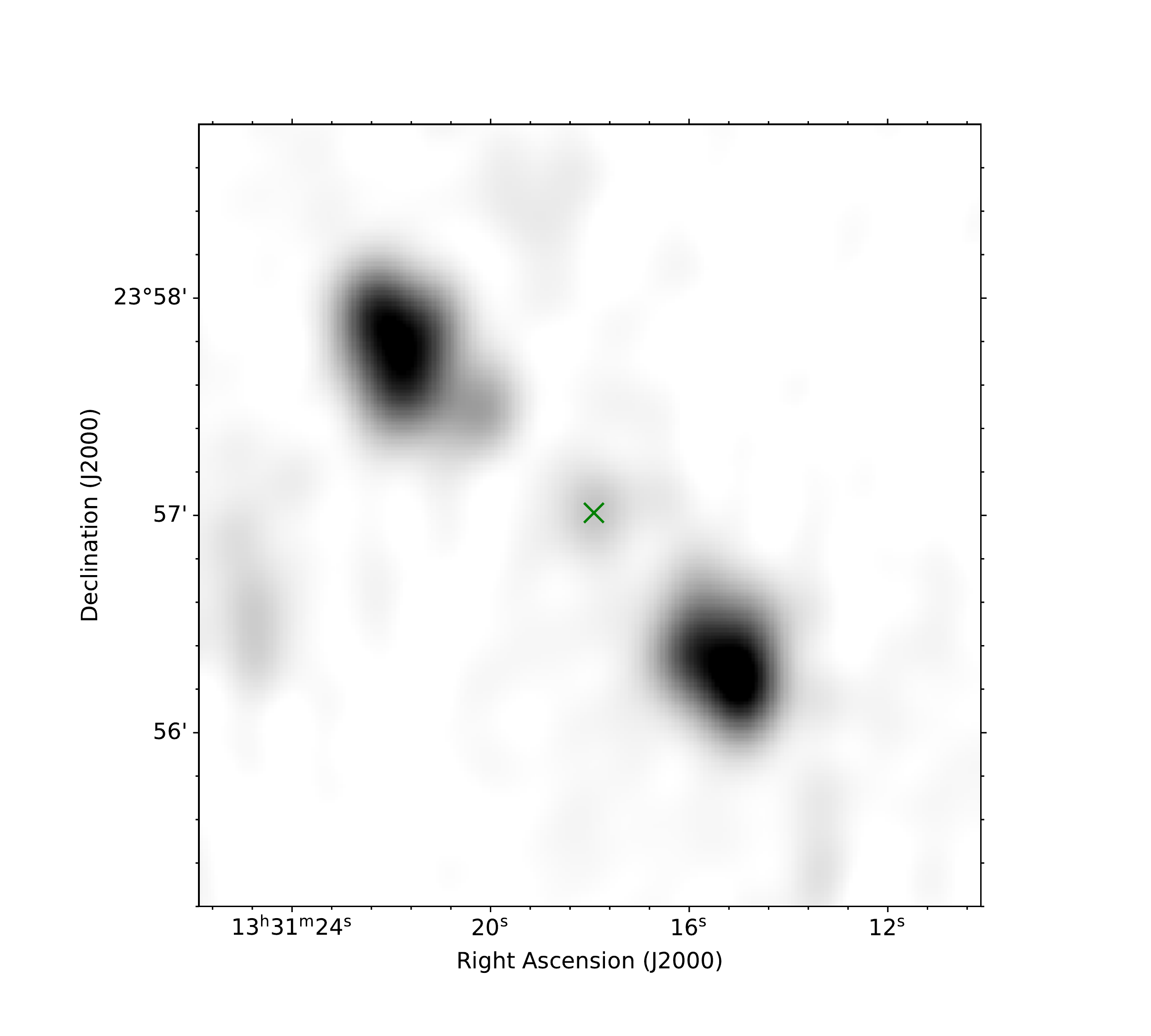}
\caption{The processed greyscale image of J1331+2357 using the VLA public project AG0635 data, where a faint but visible core is seen at the center of the image. The green cross in the image indicates the host galaxy position given in Table~\ref{tab:GRG basic information}.} 
\label{fig:AG0635}
\end{figure}

\vskip .1in
\noindent
The five GRG sources are: 
\begin{description}
\item[\textbf{J0941+3126}] This source is also known as B2\,0938+31A, and is centred at J2000.0\,RA\,9$^{\rm h}$41$^{\rm m}$01.24$^{\rm s}$~DEC\,+31$\rm ^{\circ}$26$'$32.3$''$ \citep{Colla1970,Colla1972,Colla1973,Fanti1974}. The source is hosted by SDSS J094103.62+312618.7. The source has a flux density of 20\,mJy at 15.2\,GHz \citep{Waldram2010}, and 7.2$\pm$3.3\,mJy at 30\,GHz \citep{Gawronski2010}. Its host has $W2-W3>2$, redder than is typical for elliptical galaxies and more consistent with the `intermediate disk galaxy' designation of \citet{Jarrett2017}. The host galaxy in this case currently has only photometrically determined redshifts \citep{Alam2015,Bilicki2016,Zou2019}, ranged from 0.282 to 0.398. We adopted the lowest one measured by SDSS DR12. We consequently note that this GRG candidate should be treated with caution.

\item[\textbf{J1331+2557}] This source is also known as 7C\,1328+2412, and is centred at J2000.0\,RA\,13$^{\rm h}$31$^{\rm m}$18.12$^{\rm s}$~DEC\,+23$\rm ^{\circ}$57$'$07.4$''$ \citep{Waldram1996}. Its north-east lobe is also known as TXS\,1328+242. The host galaxy of this source is identified as SDSS\,J133118.01+235700.4. We found the source has been observed at 1.4 GHz by the VLA archival public project AG0635 (Fig.~\ref{fig:AG0635}). The observation has angular resolution of 19.7 $\times$ 13.7 arcsec, along with source flux density of 172$\pm$8\,mJy. The observation shows the source has visible radio core emission cross matched with its host galaxy, and its core flux density is 6$\pm$3 mJy. Similarly to J0941+3126, the host galaxy of the source has $W2-W3>2$.

\item[\textbf{J1402+2442}] This source is also known as B2\,1400+24, and is centred at J2000.0\,RA\,14$^{\rm h}$02$^{\rm m}$25.87$^{\rm s}$~DEC\,+24$\rm ^{\circ}$41$'$53.0$''$ \citep{Colla1970,Colla1972,Colla1973,Fanti1974}. The host of this source is a close pair of galaxies, SDSS\,J140224.25+244224.3 and SDSS\,J140224.31+244226.8. The latter has a photometric redshift $z=0.299\pm0.067$ \citep{Alam2015}. We note that although we identify the  above galaxy pair as the host for this source, SDSS\,J140225.03+244218.1 is also in close proximity, see Fig.~\ref{fig:GRG_with_NVSS_contour}. This source has a photometric redshift of $z=0.208\pm0.018$ \citep{Alam2015}.

\item[\textbf{J1421+1016}] This source is also known as MRC\,1419+104, and is centred at J2000.0\,RA\,14$^{\rm h}$21$^{\rm m}$42.03$^{\rm s}$~DEC\,+10$\rm ^{\circ}$16$'$17.3$''$ \citep{Large1981,Large1991}. This source was mentioned by \citet{Amirkhanyan2015}, but not previously identified as a GRG due to differences in the estimation of both the LAS and redshift. This source has host galaxy SDSS\,J142142.68+101626.2, which is not visible in Fig.~\ref{fig:GRG_with_NVSS_contour} where we show the SDSS-i image, but can be seen clearly in \textit{WISE} 3.4\,$\mu$m data. 

\item[\textbf{J1646+3627}] The host galaxy of this source is 2MASX\,J16464260+3627107. It is the brightest cluster galaxy in the galaxy cluster GMBCG\,J251.67741+36.45295 \citep{Hao2010} and has a slightly bent morphology, see Fig.~\ref{fig:GRG_with_NVSS_contour}. This morphology is consistent with the findings of \citet{Garon2019} who used 4304 extended radio sources from RGZ to determine that BCGs have higher probabilities than other cluster members to have slightly bent morphologies. 

\end{description}

\begin{table*}
\begin{center}
\begin{tabular}{| p{1.5cm} | p{2.5cm} | p{2cm} | p{2.6cm} | p{2cm} | p{1.8cm} | p{1.5cm} |}
\hline

 \textbf{GRG} & \textbf{RGZ ID} &  \textbf{RA (J2000.0)}  & \textbf{DEC (J2000.0)}  & \textbf{z} & \textbf{LAS} & \textbf{Size} \\
   &  & \textbf{[h:m:s]}  & \textbf{[$^{o}$:$'$:$''$]}  &  & \textbf{[arcsec]} & \textbf{[kpc]} \\
\hline
J0941+3126 & J094103.6+312618 &  09:41:03.62   & +31:26:18.7 & 0.282$\pm$0.0454$^{\rm p}$ & 163 & 717 $\pm$ 88\\
J1331+2357 & J133117.9+235700 &13:31:18.01   & +23:57:00.4 & 0.33610$\pm$0.00006$^{\rm s}$ & 162  & 803 $\pm$ 7\\
J1402+2442 & J140224.3+244226 & 14:02:24.25   & +24:42:24.3 & 0.337$\pm$0.032$^{\rm p}$  & 173 & 810 $\pm$ 12\\
J1421+1016 & J142142.6+101626 & 14:21:42.68   & +10:16:26.3 & 0.37392$\pm$0.00003$^{\rm s}$ & 144 & 765 $\pm$ 6\\
J1646+3627 & J164642.5+362710 & 16:46:42.58   & +36:27:10.6 & 0.43425$\pm$0.00010$^{\rm s}$ & 130 & >754 $\pm$ 1\\
\hline 
\end{tabular}
\end{center}
\caption{A summary of the newly discovered GRGs found in the present work. RGZ ID for each source represents the truncated host galaxy coordinates recorded in the RGZ DR1 catalogue. RA/DEC of source host galaxies are that of the infrared host galaxies shown in the Fig~\ref{fig:GRG_with_NVSS_contour}. The LAS of the sources is measured using \textbf{HEALPix Ximview}. For the first four sources, we have assigned errors of five arcsecs (FWHM) to the LAS of each source, since their leading edges are fairly sharp. In the case of J1646+3627, we have listed the size as a lower limit as the source could be found to extend further given observations with improved sensitivity to larger scale structure. Redshift annotations: $\rm p$: photometric; $\rm s$: spectroscopic.}
\label{tab:GRG basic information}
\end{table*}

\setlength{\tabcolsep}{4pt}
\begin{table*}
\begin{center}
\begin{tabular}{| c | c | c | c | c |}
\hline
\textbf{GRG} & \textbf{W1}  & \textbf{W2} & \textbf{W3}  \\

\hline
J0941+3126 & 15.165$\pm$0.038  & 14.650$\pm$0.062 &  11.595$\pm$0.204\\
J1331+2357 & 14.704$\pm$0.030  & 14.441$\pm$0.048 & >12.205\\
J1402+2442 & 14.763$\pm$0.031  & 14.319$\pm$0.045 & 12.488$\pm$0.415\\
J1421+1016 & 15.104$\pm$0.033  & 14.703$\pm$0.054 & 12.841$\pm$0.512\\
J1646+3627 & 13.944$\pm$0.141  & 13.799$\pm$0.031 & >12.275\\
\hline
\end{tabular}
\end{center}
\caption{A summary of source infrared properties. \textit{WISE} magnitudes in the table are extracted from the $\rm AllWISE$ catalogue \citep{Cutri2013} via \textbf{VizieR} \citep{Vizier2000}.}
\label{tab:GRG infrared properties}
\end{table*}

\setlength{\tabcolsep}{4pt}
\begin{table*}
\begin{center}
\begin{tabular}{| c | c | c | c | c | c | c | c | c | c | c |}
\hline
\textbf{GRG} & \textbf{VLSSr} & \textbf{TGSS ADR1} & \textbf{7C} & \textbf{WENSS} & \textbf{TXS} & \textbf{NVSS} | \textbf{FIRST} & \textbf{VLASS} &  \textbf{MIT-Green} & \textbf{log\,P$_{1.4}$} \\

    & 73.8\,MHz  &147.5\,MHz & 151\,MHz &325\,MHz  & 365\,MHz    &  1.4\,GHz           & 3\,GHz   &5\,GHz       &   [W/Hz]            \\

\hline
J0941+3126  &  1160$\pm$202 & 925$\pm$68 & 930$\pm$76  &  302$\pm$87 &    &  219$\pm$26 |  144$\pm$2  &  126$\pm$1 & &  25.73   \\
J1331+2357  &  2747$\pm$676 & 1214$\pm$62 & 1270$\pm$149    &  & 901$\pm$105 &          177$\pm$22 |  100$\pm$1  &  143$\pm$1 & & 25.81   \\
J1402+2442  &  2728$\pm$541 & 1592$\pm$111&   1840$\pm$140  &        & 812$\pm$73 & 381$\pm$41 |  263$\pm$4  &  120$\pm$1 & 137 & 26.15   \\
J1421+1016  &  2078$\pm$495 & 884$\pm$67  &      &  & 701$\pm$75    &  242$\pm$30 |  184$\pm$4  &  98$\pm$1 & 106 &  26.05   \\
J1646+3627  &  521$\pm$160  & 53$\pm$6    &   &  41$\pm$15  &    &  35$\pm$6  |  30$\pm$1   &  17$\pm$3  &  & 25.36  \\
\hline
\end{tabular}
\end{center}
\caption{A summary of source radio fluxes. Fluxes in the table are measured in mJy. Surveys including the VLA Low-frequency Sky Survey Redux \citep[VLSSr;][]{Lane2014}, the NRAO VLA Sky Survey \citep[NVSS;][]{Condon1998}, FIRST \citep{Becker1995}, and VLASS \citep{Lacy2019} are done by the Karl G. Janksy Very Large Array \citep[VLA;][]{Thompson1980}. Source radio fluxes from the GMRT 150 MHz all-sky radio survey \citep[TGSS ADR1;][]{Intema2017}, the Westerbork Northern Sky Survey at 325\,MHz \citep[WENSS;][]{Rengelink1997} are also measured. We further found literature flux densities from the 7C survey of radio sources at 151\,MHz \citep{Waldram1996}, the Texas Survey of Radio Sources at 365\,MHz \citep[TXS;][]{Douglas1996} and the MIT-Green Bank Survey at 5\,GHz \citep[MG1,MG2;][]{Bennett1986,Langston1990}. Source flux densities are calibrated to a common flux scale of \citet{Scaife2012}. The radio luminosity log$\rm P_{1.4}$ is based on the NVSS images.}
\label{tab:GRG multi-survey flux summary}
\end{table*}

\section{Analysis and Discussion}
\label{sec4_0}

\begin{figure*}
\setlength{\unitlength}{1cm}
\includegraphics[width=0.9\textwidth]{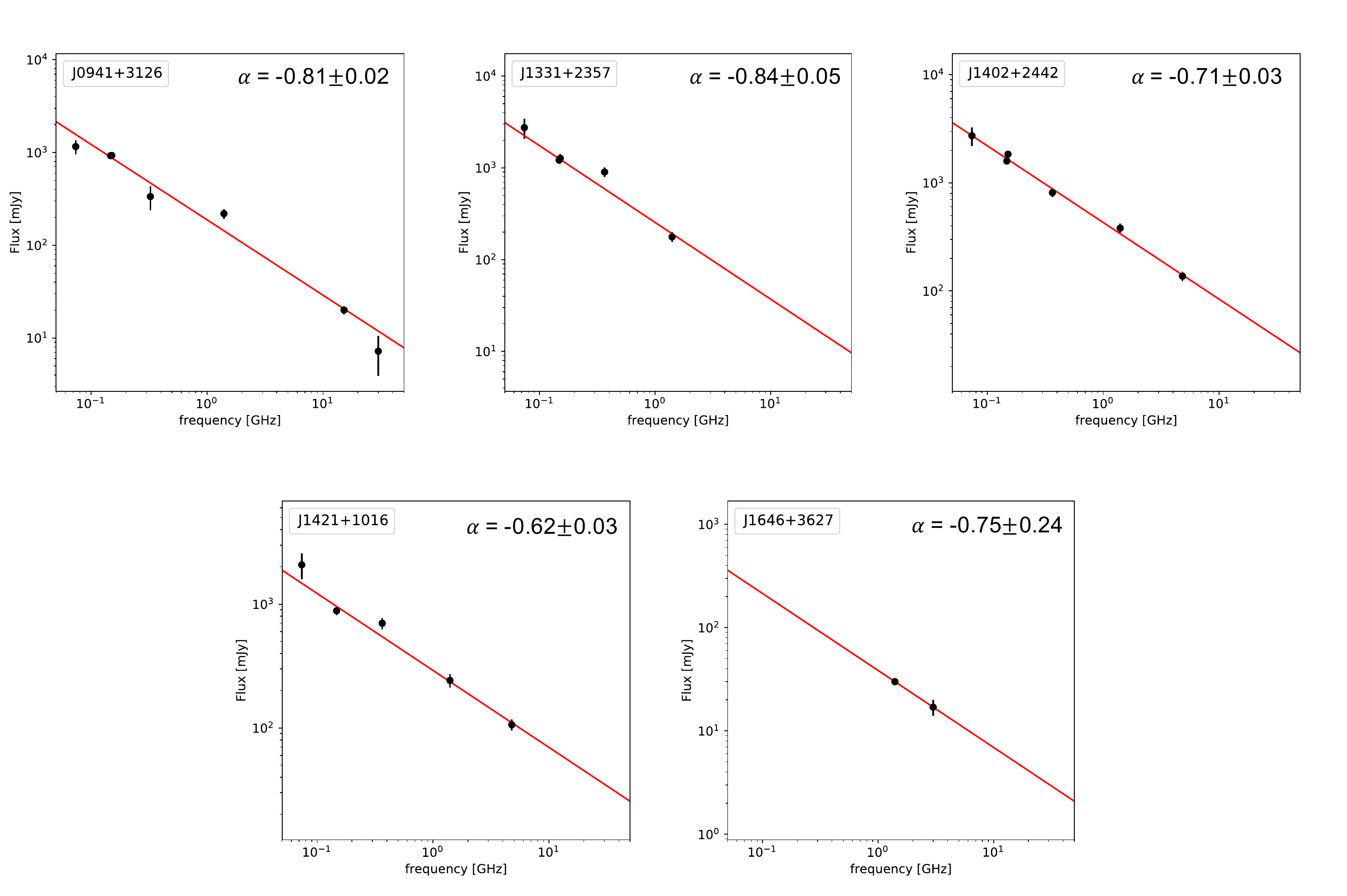}
\caption{Continuum radio spectra of the GRGs from this work. The solid red lines are linear least-squared fits, where the data points are weighted with their measurement errors when estimating the source spectral indices. Data points used for deriving source spectral indices are from Tab.\ref{tab:GRG multi-survey flux summary} and Sec.\ref{sec3_0}. Considering the angular resolution differences between surveys, we used data from the NVSS survey at 1.4\,GHz and not VLASS data for the top four sources. When deriving the source and core spectral index of J1646+3627, we consider only FIRST and VLASS as they show clear radio core emission and have comparable angular resolution. We were unable to identify clearly visible radio cores in other cited surveys.}
\label{fig:source_spectral_index}
\end{figure*}

The overall occurrence of GRGs in the RGZ~DR1 catalogue is 0.08\%, which is slightly lower than that of LoTSS DR1. There are two potential reasons for this difference. Firstly, the RGZ citizen scientists are provided with only small-sized images to classify (3$\times$3\,arcmin$^2$), which limits the LAS of radio galaxies that can be fully contained in the image cutouts. Among the 11\,237 galaxies considered in this work, the maximum source LAS is 195\,arcsec. For GRGs to have angular sizes smaller than this requires them to lie at redshifts $z\ge0.213$. Under a similar restriction, \citet{Dabhade2019} would have missed 26.3\% of their discovered GRGs. Correspondingly, the \citet{Kozie2020} and \citet{Kuzmicz2018} samples would have missed as much as 62.5\% and 66.4\% of their catalogued GRGs within the sky area covered by RGZ DR1, respectively. We also take the redshift limitation given by SDSS DR12 into account, while SDSS DR12 is able to detect quasar as far as $z=6.440$ \citep{Alam2015}, which is further than any detected GRG. Secondly, GRGs have historically been poorly detected in radio surveys like FIRST in part due to their synchrotron spectral index. The radio lobes of GRGs share relatively steep spectral indices, i.e. they are brighter at lower frequencies and thus in principle can more easily be found at MHz frequencies compared to GHz \citep{Dabhade2019}. 

In addition, finding GRGs in radio surveys like FIRST is limited by instrumental considerations. Interferometers with comparatively long baselines (as a function of wavelength) may not be sensitive to the large-scale emission associated with extended or diffuse radio sources \citep{Saxena2018}, nor may it always be encompassed by the comparatively small field-of-view for single-pixel centimetre-wave receivers. Such issues have in part been alleviated by radio telescopes such as the Expanded Very Large Array \citep[EVLA;][]{Sahr2002}, and LOFAR at MHz-frequencies, and by telescopes with large instantaneous fields of view due to Phased Array Feed (PAF) technology, such as the Australian SKA Pathfinder \cite[ASKAP;][]{Johnston2008}; however, whilst these instruments may be powerful probes of GRGs in the future \citep{Peng2015} instrumental selection effects will always persist. 

Consideration of selection effects is of particular importance in the context of developing automated deep learning based GRG classifiers. Such algorithm development is complicated by a lack of large, uniform, and reliable cross-matched radio source catalogues that contain source information characterised in a consistent manner appropriate for the formation of computationally tractable training data. Furthermore, a key aspect of the development of potential machine learning based GRG classification algorithms, as well as radio galaxy classification more generally, is a clear understanding of the biases that are introduced by this training data selection. In this respect the RGZ DR1 catalogue represents a well-understood data sample where considerations such as input image size are pre-defined. Hence, although the restricted image size is considered a disadvantage for compiling large catalogues of GRGs, it is potentially an advantage for defining a deep learning training dataset with well understood data constraints.

\subsection{Radio Source Luminosity}
\label{sec4_1_prime}

We measured the integrated flux densities for each source using images from the VLSSr, TGSS ADR1, WENSS, NVSS, FIRST and VLASS surveys. We also retrieved literature integrated source flux densities from the 7C, TXS, 9C, and MIT-Green Bank (MG) surveys using the NASA/IPAC Extragalactic Database\footnote{NED is operated by the Jet Propulsion Laboratory, California Institute of Technology,
under contract with the National Aeronautics and Space Administration.}, these are listed in Table~\ref{tab:GRG multi-survey flux summary}. At low frequencies, all historic data in Table~\ref{tab:GRG multi-survey flux summary} has been re-scaled to match the \citet{Scaife2012} flux density scale, which is consistent with the \citet{Perley2017} flux scale at higher frequencies. The five GRGs identified in this paper have higher integrated flux densities in the NVSS survey (FWHM = 45$''$) than the FIRST survey (FWHM = 5.4$''$), which is consistent with a lack of shorter baseline coverage in the FIRST survey compared to that of NVSS. We note that the VLASS measurement should also be treated with caution as all five objects have diffuse emission on angular scales larger than 30$''$, which will be poorly recovered by this survey and result in underestimated integrated flux densities \citep{Lacy2019}.

The resulting spectra for all sources are shown in Fig.~\ref{fig:source_spectral_index}. We find that these GRGs have a range of source spectral indices from $-0.84< \alpha < -0.62$, with an average spectral index of $\langle \alpha \rangle = -0.75$. This is similar to the mean spectral index, $\langle \alpha \rangle_{0.151}^{1.4} = -0.79$, found for the GRG sample of \citet{Dabhade2019} and is also consistent with the typical value for radio galaxies more generally \citep[e.g.][]{Kuzmicz2018}. Finally, our result happens to have the same view with \citet{Hardcastle2019,Shabala2020} that these long-lived large radio galaxies are the tail of the radio galaxy age distribution.

Since FIRST and VLASS have comparable resolution and flux loss on similar scales, we also compute the spectral index, $\alpha_{1.4}^{3.0}$, of the source core and lobes separately for J1646+3627 where the radio core is visible and has peak flux above 3$\rm \sigma_{rms}$ in both surveys. We found that the core region of the source has $\rm S_{1.4}=2.99\pm1.21\,mJy$ and $\rm S_{3}=2.7\pm0.3\,mJy$, giving a source core spectral index of $\alpha_{1.4}^{3.0} = -0.13$, and $\alpha_{1.4}^{3.0} < -0.69$ for the lobes. This is consistent with other resolved radio systems, where super-position of multiple synchrotron emission components result in a flatter core spectrum. 

\subsection{GRGs that are also BCGs}
\label{sec4_1}

The GRG J1646+3627, newly identified here, is also the brightest cluster galaxy in the galaxy cluster $\rm GMBCG\,J251.67741+36.45295$ \citep{Hao2010}.  To better understand this emerging population, we performed a literature search for GRGs that are already known but have not previously been identified as a BCG. Following \citet{Dabhade2019}, we cross-matched the \citet{Kuzmicz2018} catalogue with the GMBCG \citep{Hao2010} and WHL \citep{Wen2012} galaxy cluster catalogues. 

This search returned 13 new BCG GRG candidates, which are listed in Table~\ref{tab:BCG_GRG}. 10 out of 13 of these candidates have not been identified as BCGs previously due to a historic lack of availability of large scale optical galaxy cluster catalogues when they are discovered. The other 3 candidates are not recognized as finding BCG GRGs was not highlighted in \citet{Proctor2016}.

\begin{figure}
\setlength{\unitlength}{1cm}
\includegraphics[width=0.45\textwidth]{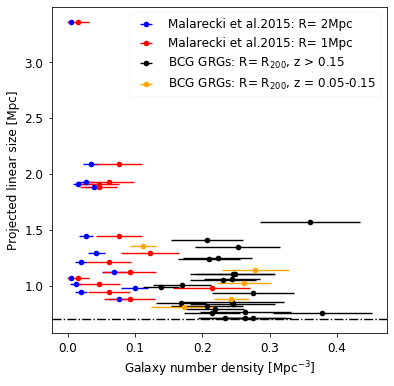}
\caption{A diagram of galaxy number density vs. source projected linear size, comparing samples discussed in \citet{Malarecki2015} and BCG GRGs with $\rm R_{200}$ and $\rm N_{200}$ available from the WHL catalogue. $\rm R_{200}$ and $\rm N_{200}$ of each BCG GRGs in the diagram can be found in Table~\ref{tab:BCG_GRG} and the Table 3 of \citet{Dabhade2019}. The galaxy number density uncertainty of the BCG GRGs are estimated based on the Equation 1 of \citet{Wen2012} and our cylindrical volume assumption. The galaxy number density uncertainty of \citet{Malarecki2015} samples are extracted from the Table 4 of their work. The dashed line in the diagram equals 700 kpc.}
\label{fig:BCG_num_density_vs_linear_size}
\end{figure}

Prior to this work, 21 BCG GRGs were identified by  \cite{Dabhade2017,Dabhade2019}. Combining that sample with the 13 we identify from the literature and the 1 new GRG BCG from RGZ DR1, there are 35 BCG GRGs known in total. From the full sample of 35 BCG GRGs, 28 are in clusters with catalogued $R_{200}$, the radius where the mean density is at least 200 times the critical density of the Universe, and $N_{200}$, the local galaxy number within $R_{200}$ \citep{Wen2012}. $N_{200}$ here only counts galaxies with $M_{\rm e}^{r}(z) \le -20.5$, where $M_{\rm e}^{r}(z)$ refers to evolution-corrected absolute magnitude in the $r$ band \citep{Wen2012}: $ M_{\rm e}^{r}(z) = M_{r}(z) + Qz$ , for which a passive evolution of $Q=1.62$ was adopted \citep{Blanton2003}.
Using these data, the relationship between local galaxy density and projected linear size for these galaxies is shown in Fig.~\ref{fig:BCG_num_density_vs_linear_size}. Of these 28 objects, there are five with host galaxy redshifts $0.05<z<0.15$, the same range used by \citet{Malarecki2015} who also investigated the relationship between  GRG size and local environment. We re-calculate the galaxy number density of each cluster for these five objects assuming a cylindrical volume with a radius of $\rm R_{200}$ for each source. Consistent with \citet{Malarecki2015}, we adopt a physical cylinder length equivalent to $z=0.1\pm0.003$. This returns galaxy number density values from 0.11 to 0.27\,$\rm Mpc^{-3}$ with a median galaxy number density of 0.24\,$\rm Mpc^{-3}$. These are shown as yellow data points in Fig.~\ref{fig:BCG_num_density_vs_linear_size}. Original data from \citet{Malarecki2015} are shown as blue data points; however, the values of $R_{200}$ for these galaxies are generally closer to 1\,Mpc than the cylinder radius of 2\,Mpc used by Malarecki et~al., consequently we also show the local galaxy density for the sources in \citet{Malarecki2015} re-calculated using a cylinder radius of 1\,Mpc. These data are shown as red points in  Fig.~\ref{fig:BCG_num_density_vs_linear_size}. The maximum galaxy number density of the 23 BCG GRGs with with host galaxy redshifts $z>0.15$ under the same volume assumption is 0.38\,$\rm Mpc^{-3}$.

From Fig.~\ref{fig:BCG_num_density_vs_linear_size} it can be seen that the BCG GRGs have been growing in generally denser environments than the non-cluster/poor cluster GRGs in the sample of \citet{Malarecki2015}. When considering a radius of 1\,Mpc, source B\,1308-441 from the  \citet{Malarecki2015} sample has a  comparable local galaxy number density to the BCG GRGs, due to a concentration of galaxies in close proximity. The mean galaxy number density of the \citet{Malarecki2015} GRGs using a cylinder radius of 1\,Mpc is 0.07\,$\rm Mpc^{-3}$, typical for a poor cluster or galaxy group. 

For the BCG GRGs we also compute the cluster mass, $\rm M_{200}$, in each case. With the exception of WHL\,J112126.4+534457 with a mass of 4.6$\times 10^{14}\rm M_{\odot}$, the masses of these clusters lie in the range $0.7-2\times 10^{14}\,\rm M_{\odot}$. Given that the average $\rm M_{200}$ for the WHL catalogue is $\sim$1.12$\times 10^{14}\,\rm M_{\odot}$, the masses of these particular clusters are unremarkable with respect to the wider catalogue.

\begin{table*}
\begin{center}
\begin{tabular}{| p{1.8cm} | p{4.2cm} | p{1.5cm} | p{1.5cm} | p{1cm} | p{1cm} | p{1cm} | p{1cm} | p{1cm} | p{1cm} | p{1cm} |}
\hline

 \textbf{GRG ID} & \textbf{Cluster ID}  & \textbf{RA (J2000.0)}  & \textbf{DEC (J2000.0)}  & \textbf{z} & \textbf{$\rm R_{200}$} & \textbf{$\rm N_{200}$} & \textbf{$\rm R_{L*}$} & \textbf{$\rm M_{200}$} & \textbf{FR type} & \textbf{Ref.}\\
   &  & \textbf{[h:m:s]}  & \textbf{[$^{o}$:$'$:$''$]}  &  & \textbf{[Mpc]}  &  &  & $\rm [10^{14} M_{\odot}]$  &\\
\hline
J1054+0227	     &GMBCG J163.58817+02.46528 & 10:54:21.16   & +02:27:55.0    &0.34   & $-$   & $-$ & $-$   & $-$ &II   &7\\
J1400+3019	     &GMBCG J210.18097+30.32185	& 14:00:43.43	& +30:19:18.7    &0.206	& $-$   & $-$ & $-$	  & $-$ &II   &6\\
J0115+2507	     &WHL J011557.2+250720	    & 01:15:57.23	& +25:07:21.0	&0.1836	& 0.96	& 15  & 18.28 & 1.0 &II   &7\\
J0129$-$0758 &WHL J012935.3$-$075804 & 01:29:35.26 & $-$07:58:04.3	&0.0991$^{a}$	& 1.17	& 10  & 28.44 & 1.6 &I/II &4\\
J0751+4231	     &WHL J075108.8+423124	    & 07:51:08.80 & +42:31:24.2	&0.2042	& 0.98	& 14  & 17.87 & 0.9 &II   &8\\
J0902+1737 &WHL J090238.4+173751	    & 09:02:38.42 & +17:37:51.5	    &0.1645$^{a}$	& 1.01	& 14  & 19.68 & 1.1 &II   &7\\
J0926+6519	     &WHL J092600.8+651923	    & 09:26:00.82 & +65:19:22.7	    &0.1397	& 0.84	& 8   & 14.41 & 0.7 &I 	  &3\\
J1108+0202 &WHL J110845.5+020241	    & 11:08:45.49	& +02:02:40.9	    &0.1574$^{a}$	& 1.05	& 26  & 23.55 & 1.3 &II   &4\\
J1235+2120	     &WHL J123526.7+212035	    & 12:35:26.67	& +21:20:34.8	    &0.4227	& 0.79	& 10  & 12.03 & 0.6 &II   &5\\
J1418+3746	     &WHL J141837.7+374625	    & 14:18:37.65 & +37:46:24.5	&0.1349	& 1.17	& 25  & 28.14 & 1.6 &II   &8\\
J1453+3308	     &WHL J145302.9+330842	    & 14:53:02.86	& +33:08:42.4	&0.2482	& 0.92	& 14  & 16.69 & 0.9 &II   &8\\
J1511+0751	     &WHL J151100.0+075150	    & 15:11:00.01	& +07:51:50.0	    &0.4594	& 1.09	& 17  & 23.20 & 1.3 &II   &1\\
J2306$-$0930	 &WHL J230632.2$-$093020 	    & 23:06:32.18	& $-$09:30:20.6	&0.1593	& 1.03	& 16  & 20.35 & 1.1 &I	  &2\\
\hline \hline
\end{tabular}
\end{center}
\caption{A summary of the BCG GRG candidates we found from \citet{Kuzmicz2018}. RA/DEC, redshift, FR type, and Reference number are extracted from \citet{Kuzmicz2018}. The galaxy cluster ID are extracted from GMBCG \citep{Hao2010} and WHL \citep{Wen2012} galaxy cluster catalogues. $\rm R_{200}$: the radius of a cluster that its mean density is 200 times of the critical density of the universe; $\rm N_{200}$: the galaxy number within the $\rm R_{200}$; $\rm R_{L*}$: cluster richness; $\rm M_{200}$: the mass of a cluster that its mean density is 200 times of the critical density of the universe, which is derived from $\rm R_{L*}$ using the Equation 2 of \citet{Wen2012}. \hspace{\textwidth}\textbf{References}: 1. \citet{Baum1989}, 2. \citet{Best2005}, 3. \citet{Lara2001b}, 4. \citet{Machalski2007}, 5. \citet{Nilsson1998}, 6. \citet{Parma1996},  7. \citet{Proctor2016}, 8. \citet{Schoenmakers2001}.}\hspace{\textwidth}$^{a}$: The cluster redshift and the source redshift have a difference of $0.03-0.04$, the cluster membership of these radio sources should be treated which caution.
\label{tab:BCG_GRG}
\end{table*}

\section{Conclusion}
\label{sec5_0}

In this work we have identified 5 new GRGs from RGZ DR1. These GRGs mostly share an FR~II radio morphology and cover the redshift range of $0.28<z<0.43$. These GRGs have been identified using a method consistent with the assembly of training data appropriate for a deep learning classifier. We compare the selection of these GRGs to previous studies and suggest that samples defined in this manner are more likely to be representative of future deep learning approaches to GRG identification than previous methods.

We associate one of the newly identified GRGs, J1646+3627 with the brightest cluster galaxy in galaxy cluster GMBCG\,J251.67741+36.45295 \citep{Hao2010} and using literature data we identify a further 13 previously known GRGs to be BCG candidates. This increases the number of known BCG GRGs by more than 60\%. We show that the local galaxy density of these sources is significantly higher than that of non-cluster GRGs, challenging the hypothesis that GRGs are able to grow to such large sizes only due to locally under-dense environments.

\section*{Acknowledgements}

The authors are grateful for the contributions of over 12,000 volunteers in the Radio Galaxy Zoo project, who are acknowledged at \url{http://rgzauthors.galaxyzoo.org}. In particular we note the RGZ users \emph{antikodon}, \emph{Dolorous\_Edd}, and \emph{WizardHowl}, who made notes on  4 out of 5 galaxies discussed in this paper in the RadioTalk forum. The authors are also very grateful for discussions from the machine learning group at Jodrell Bank Centre for Astrophysics, JBCA. This research was supported by JBCA, University of Manchester. The corresponding author is also grateful for the effort of Katie Hesterly, Alex Clarke, Emma Alexander, and project team members, participating school teachers and students of RGZ\_CN, a teaching side project of the Radio Galaxy Zoo. Partial support for the work of LR comes from NSF grant AST17-14205 to the University of Minnesota. Partial support for the work of AK comes from the National Radio Astronomy Observatory, a facility of the National Science Foundation operated under cooperative agreement by Associated Universities, Inc. AMS gratefully acknowledges support from the Alan Turing Institute.








\appendix




\bsp	
\label{lastpage}
\end{document}